\documentclass[12pt]{article}
\usepackage[left=.7in,right=0.7in,top=1in,bottom=1in,footnotesep=.25in]{geometry}
\usepackage{authblk}
\usepackage[margin=1cm]{caption}

\usepackage{amssymb}
\usepackage{amsmath}
\usepackage{xcolor}
\usepackage{graphicx}
 \usepackage{floatrow}
 \usepackage{soul}

\def\be{\begin{equation}}
\def\ee{\end{equation}}
\def\bea{\begin{eqnarray}}
\def\eea{\end{eqnarray}}
\def\ba{\begin{array}}
\def\ea{\end{array}}
\def\bse{\begin{subequations}}
\def\ese{\end{subequations}}

\def\pd{{\partial}}

\def\p{{\varphi}}

\date{}
\title{Formation of caustics in k-essence \\
and  Horndeski theory}

\author{Eugeny~Babichev$^1$} 
\affil{$^1$Laboratoire de Physique Th\'eorique, CNRS, Univ. Paris-Sud, \\ Universit\'e Paris-Saclay, 91405 Orsay, France}

\begin{document}

\maketitle

\abstract{
We study propagation of waves and appearance of caustics in k-essence and galileon theories.
First we show that previously known solutions for travelling waves in k-essence and galileon models correspond to very specific fine-tuned initial conditions.
On the contrary, as we demonstrate by the method of characteristics, generic initial conditions leads to a wave in k-essence which ends up with formation of caustics. 
Finally, we find that any wave solution in pure k-essence is also a solution for a galileon theory with the same k-essence term. 
Thus in the Horndeski theory with a k-essence term formation of caustics is generic. We discuss physical consequences of the caustics formation and possible ways to cure the problem.}


\section{Introduction}
Scalar fields play a crucial role in modern cosmology and modified gravity. 
The early and present day accelerated expansion  of the Universe are often attributed to the presence of a scalar field, 
which drives the acceleration. 
An introduction of a scalar field is also a simple and natural way to build modify gravity models.

The simplest scalar field model is a canonical massive scalar field, described by the Lagrangian\footnote{We use ``mostly positive'' signature convention $+---$ throughout the paper.}, 
\begin{equation}\label{Lcan}
\mathcal{L} = \frac12(\partial\phi)^2 - \frac12 m^2 \phi^2.
\end{equation}
A simple generalisation of the above Lagrangian is to replace the standard mass term by an arbitrary function, i.e. to consider the following Lagrangian,
\begin{equation}\label{Lquint}
\mathcal{L} = \frac12(\partial\phi)^2 - V(\phi).
\end{equation}
This type of Lagrangian has been popular in the literature for some time by the name quintessence as a model to explain Dark energy~\cite{Ratra:1987rm}. 

A further generalization of the scalar field Lagrangian leads to so-called k-essence models, whose Lagrangian is an arbitrary function of 
the standard kinetic term 
$X\equiv \frac12(\partial_\mu\phi\partial^\mu\phi)$ and the field itself,
\begin{equation}
\mathcal{L}= \mathcal{L}(X,\phi) \label{Lk}
\end{equation}
In the cosmological context this type of Lagrangian has been suggested in \cite{ArmendarizPicon:1999rj} as an inflation model and in \cite{ArmendarizPicon:2000ah} 
as a Dark energy model.
The k-essence Lagrangian leads to second-order quasilinear equation of motion and no dangerous Ostrogradski ghost is present in the theory, 
since only up to second-order time derivatives are present in the equations of motion (see a relevant discussion, e.g. in~\cite{Woodard:2006nt}). 

The k-essence model is not the most general scalar field Lagrangian, leading to the second order equation of motion. 
In fact the most general Lagrangian, yielding second-order equations of motion (both in metric and the scalar field) are found by Horndeski in \cite{Horndeski} 
and then rediscovered later in other works \cite{Nicolis:2008in,Deffayet:2009wt,Deffayet:2011gz}.
The Horndeski Lagrangian contains, apart from the first derivatives, also the second derivatives of the scalar field, i.e. it has a form 
\begin{equation}
	\mathcal{L}=H(\phi, X, \nabla\nabla\phi),
\end{equation}
where the precise expression for $H$ is given later in the paper\footnote{
Note that although usually higher-order equations of motion mean the presence of the Ostrogradski ghost, 
as happens for example for a theory $\mathcal{L} \sim (\Box\phi)^2$ (see~\cite{Anisimov:2005ne} for cosmological applications), 
there are theories beyond Horndeski, which do contain higher-order derivatives in the equations of motion, but nevertheless do not contain 
extra degrees of freedom~\cite{Gleyzes:2014dya,Gleyzes:2014qga,Lin:2014jga,Deffayet:2015qwa,Langlois:2015cwa} (see also~\cite{Gao:2014soa} and~\cite{Chamseddine:2013kea}).
}.

The k-essence model, and more generically --- the Horndeski theory, has a peculiar property: 
the speed of propagation of the scalar field perturbations does not coincide with the speed of gravity. The reason is the non-linearity of the equations of motion.
Indeed, for the canonical~(\ref{Lcan}) or the quintessence model~(\ref{Lquint}), the speed of propagation of perturbations coincides with the speed of gravity, because the equations of motion are linear. 
On the contrary, for the k-essence or the galileon, the kinetic matrix for perturbations depends on the background solution, and therefore in general the speed of propagation is not equal to 1. 
It should be noted that both sub- and super-liminal propagation of perturbations may exist, depending on the model and the background solution. 
As it was argued in \cite{Bruneton:2006gf,Babichev:2007dw}, 
this does not necessarily lead to the acausality problem in k-essence. Similar arguments can be given for the galileon model~\cite{Burrage:2011cr}.

In~\cite{Babichev:2007dw} it was noted that for any non-linear pure k-essence model, $\mathcal{L}=\mathcal{L}(X)$,
there are always two separate plane wave solutions, 
\begin{equation}\label{trav}
\phi = f_1(t+x), \;\;\; \phi=f_2(t-x),
\end{equation}
with $f_1$ and $f_2$ being arbitrary. 
This result has been obtained under the assumption that the metric is Minkowski and non-dynamical.
Note that each of these solutions is also a solution for the canonical scalar field $\mathcal{L}=\frac12(\partial\phi)^2$, and moreover, 
the general solution in $1+1$ dimensions for the canonical scalar field is the sum of $f_1$ and $f_2$.
Due to the non-linearity, $f_1+f_2$ is not a solution for k-essence in general. 

A similar result holds also for the galileon field, as it has been shown in \cite{Evslin:2011vh,Evslin:2011rj}: 
a subclass of the galileon model supports travelling waves (on fixed Minkowski spacetime), 
i.e. solutions of the form~(\ref{trav}). 
This result has been later generalised to the case of the most general Horndeski theory with a dynamical metric~\cite{Babichev:2012qs}.

In this paper we study in detail propagation of waves in the the shift-symmetric k-essence and galileon models in 1+1 dimension spacetime.
We first show that in the case of the k-essence theory~(\ref{Lk}), the travelling wave, Eq.~(\ref{trav}), is not a generic solution and it corresponds to 
very particular initial conditions. 
On the contrary, as we demonstrate by using the method of characteristics, 
a generic wave solution in k-essence does not keep its form when propagating
and eventually leads to the formation of caustics\footnote{In~\cite{Felder:2002sv} 
it was shown that caustics are formed in a specific non-canonical scalar field model, 
the Born-Infeld field theory, due to the fact that there is a regime for which the Born-Infeld scalar field behaves 
as dust, see also \cite{Goswami:2010rs}. 
Here, however, we show that the caustics is a generic feature for all k-essence models, independently of whether they behave as dust in some regime or not.
Other theories, which are known to have the problem of caustics include ghost condensate~\cite{ArkaniHamed:2005gu} 
and the Gauss-Bonnet theory~\cite{ChoquetBruhat:1988dw} (see discussion e.g. in~\cite{Deruelle:2003ck}).} 
Finally, any wave solution for the pure k-essence model $\mathcal{L}=K(X)$ in $1+1$ dimensional spacetime 
is also a solution for the shift-symmetric galileon model $\mathcal{L} = K(X) + H(X,\nabla\nabla\phi)$, where $H(X,\nabla\nabla\phi)$
is the most general shift-symmetric Galileon Lagrangian containing second derivatives\footnote{
With this requirement we exclude the k-essence term from $H(X,\nabla\nabla\phi)$, but allow all other shift-symmetric galileon terms. 
So that in the Lagrangian the k-essence term is presented solely by $K(X)$.}. 
We discuss the obtained results and outline open issues.

\section{Dynamics in two dimensions and characteristics}
A general action for a shift-symmetric scalar field k-essence reads,
\be
\label{L}
S_K = \int d^4x \sqrt{-g} \mathcal{L}(X) ,
\ee
where $X\equiv \frac12(\pd_\mu\p\pd^\mu\p)$ is the canonical kinetic term of a scalar field. Eq.~(\ref{L}) is invariant under the 
transformation $\p \to \p +$const, hence the name ``shift-symmetric''.
The simplest example of the above k-essence action is the standard massless scalar field with $\mathcal{L} (X) = X$.
In addition, through the main part of the paper we assume that the metric is flat and non-dynamical, 
therefore the only dynamical variable in the theory is the scalar field.
In this case the equations of motion for the scalar field is linear, while for a generic $\mathcal{L}(X)$ the equations of motion is nonlinear.
Variation of (\ref{L}) with respect to $\p$ gives equation of motion (see e.g. Ref.~\cite{Rendall:2005fv,Babichev:2007dw,Goulart:2011rs}),
\begin{equation}
\label{eom0}
	\left( \mathcal{L}_X g^{\mu\nu}+\mathcal{L}_{XX}\nabla^\mu\p\nabla^\nu\p \right)\nabla_\mu\nabla_\nu\p 
	=0,
\end{equation}
where the subscript denotes the corresponding derivative, i.e. $\mathcal{L}_X\equiv d\mathcal{L}/dX$,  $\mathcal{L}_{XX}\equiv d^2\mathcal{L}/(dX)^2$.

To simplify the study, we restrict ourselves to the case of two-dimensional motion, i.e. 
$\p$ is a function of the time coordinate $t$ and one spatial coordinate $x$. 
Taking into account that the equation of motion (\ref{eom0}) does not depend on $\p$ explicitly, 
it will be convenient to define the following variables, 
\begin{equation}\label{defp}
	\tau = \dot \phi, \ \ \ \chi = \phi',
\end{equation}
where dot denotes derivative with respect to time and prime is the derivative with respect to $x$. 
The consistency $\frac{d}{dt}(\frac{d}{dx}\phi)=\frac{d}{dx}(\frac{d}{dt}\phi)$ requires the following relation to be hold,
\begin{equation}\label{cons}
\tau'=\dot \chi.
\end{equation}
In terms of the new variables (\ref{defp}) the kinetic term reads, $X=\frac12(\tau^2-\chi^2)$.
Using (\ref{defp}) and (\ref{cons}),  Eq.~(\ref{eom0}) can be rewritten in the following form,
\begin{equation}\label{eom1}
	A \dot\tau  + 2 B  \tau' + C \chi' =0,
\end{equation}
where we defined, 
\begin{equation}
\label{def}
A = \mathcal{L}_X +\tau^2 \mathcal{L}_{XX},\;\;
 B = -\tau \chi \mathcal{L}_{XX} ,\;\;
 C = -  \mathcal{L}_X + \chi^2 \mathcal{L}_{XX}.
\end{equation}
It is easy to see that in the case of the canonical scalar field, $\mathcal{L} = X$, the above equation takes the form  
$\dot\tau   - \chi' =0$, which by substitution of (\ref{defp}) assumes the form of the wave equation 
$\ddot\p - \p''=0$.
In general, however, the coefficients $A,B,C$ are functions of $\tau$ and $\chi$.

We will study (\ref{eom1}) by the method of characteristics\footnote{In this section we mostly follow the mathematical literature on
quasilinear differential equations, see e.g.~\cite{Courant} and \cite{Vladimirov}.}.  
First, let us consider an arbitrary smooth curve in the $(x,t)$ plane, with a parameter $\sigma$ along the curve. 
The derivatives of the coordinates $t$ and $x$ along the curve we denoted by $t_\sigma\equiv dt/d\sigma$ and $x_\sigma \equiv dx/d\sigma$.
Then we can easily compute the derivatives of $\tau$ and $\chi$ along the curve $\sigma$ in terms of $x_\sigma$ and $t_\sigma$,
\begin{equation}
\label{pipsi}
\begin{split}
	\tau_\sigma &= \dot \tau t_\sigma + \tau' x_\sigma,\\
	\chi_\sigma &= \dot \chi t_\sigma + \chi' x_\sigma.
\end{split}
\end{equation}
Using (\ref{pipsi}) (and assuming non-zero $t_\sigma$  and $x_\sigma$ along the curve $\sigma$) 
Eq.~(\ref{eom1}) can be rewritten as follows,
\begin{equation}\label{eom1bis}
 	\frac{A}{t_\sigma} \tau_\sigma + \frac{C}{x_\sigma}\chi_\sigma - \frac{\tau'}{\xi}\left(A\xi^2 -2B\xi +C \right)=0,
\end{equation}
where we introduced the derivative along the curve $\xi \equiv (dx/dt)_\sigma = x_\sigma/t_\sigma$. 
Now, if the expression in the parentheses vanishes, 
\begin{equation}\label{char}
	A \xi^2 -2 B\xi + C =0,
\end{equation}
then Eq.~(\ref{eom1bis}) becomes an ordinary (in general nonlinear) differential equation,
$$
(\xi A)d\tau+Cd\chi=0,
$$
which holds along the curve $\sigma$.
Eq.~(\ref{char}) is called the characteristic equation, and the solutions of the characteristic equations $\xi$ 
are the characteristics. 
Physically, the characteristic curve describes the propagation of a signal, made of small perturbations, on top of a particular solution. 
Note that the signal speed does not coincide in general neither with the group velocity nor phase velocity,
since the characteristic curve corresponds to the high-frequency limit $\omega\to \infty$.
See a relevant discussion, e.g. in~\cite{Babichev:2007dw}.
Provided that 
\begin{equation*}
B^2 - AC>0,
\end{equation*}
which, by use of (\ref{def}) reads,
\begin{equation}\label{hyper}
	\mathcal{L}_X^2 + 2X \mathcal{L}_X \mathcal{L}_{XX} >0
\end{equation}
the characteristic equation~(\ref{char}) has two real roots $\xi = \xi_{\pm}$, giving two families of characteristics,
\begin{equation}\label{xipm}
	\xi_{\pm}= \frac{-\tau\chi \mathcal{L}_{XX}\pm  \sqrt{\mathcal{L}_X^2+2X\mathcal{L}_X\mathcal{L}_{XX}}}{\mathcal{L}_X+\tau^2 \mathcal{L}_{XX}}.
\end{equation}
Notice that the condition~(\ref{hyper}) coincides with the hyperbolicity condition for the k-essence field, 
see e.g. Ref.~\cite{Babichev:2007dw}. When the expression in the square root is positive, i.e. the characteristic equation has two real roots, 
the equation is hyperbolic, while two complex roots correspond to an elliptic equation. 

Thus, provided that the hyperbolicity condition (\ref{hyper}) is satisfied, 
the partial differential equation~(\ref{eom1}) is now rewritten as 
a system of four ODEs on $t$, $x$, $\tau$ and $\chi$ as functions of two independent variables $\sigma_+$ and $\sigma_-$,
\begin{eqnarray}
	&& \frac{dx}{d\sigma_+} = \xi_+ \frac{dt}{d\sigma_+},\ \ 	\frac{dx}{d\sigma_-} = \xi_- \frac{dt}{d\sigma_-},   \label{char1}\\
	 && \left(\xi_+A\right) \frac{d\tau}{d\sigma_+} + C \frac{d\chi}{d\sigma_+} =0, \ \  
	 \left(\xi_-A\right) \frac{d\tau}{d\sigma_-} + C \frac{d\chi}{d\sigma_-} =0,\label{char2} 
\end{eqnarray}
where $\sigma_+$ and $\sigma_-$ are parameters along the characteristics $\xi_+$ and $\xi_-$ correspondingly,
Eq.~(\ref{char1}) is simply the definition of characteristics, and Eq.~(\ref{char2}) is (\ref{eom1bis}) with (\ref{char}) taken into account.
It is still a complicated problem to analyse the system (\ref{char1}) and (\ref{char2}), but luckily, in our case of a shift-symmetric k-essence field, 
the problem is simplified. Indeed, note that the equations on $\tau$ and $\chi$, i.e. Eq.~(\ref{char2}),
decouple from the other two equations, Eq.~(\ref{char1}), 
since $t$ and $x$ do not enter Eq.~(\ref{char2}) explicitly. 
Such set of equations is called {\it reducible} system~\cite{Courant}. 

\begin{figure}[t]
\includegraphics[width=0.85\textwidth]{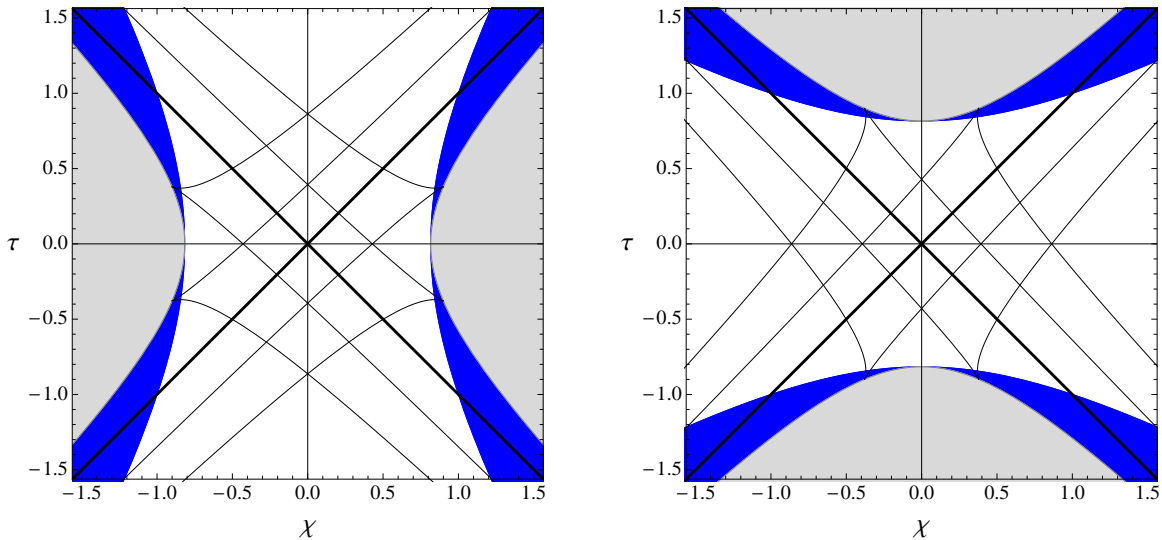}}{\caption{Image of characteristics in $(\tau,\chi)$ plane for the Lagrangian
 	$ \mathcal{L}(X) = X+ \frac12 X^2 $ (left plot) and $ \mathcal{L}(X) = X - \frac12 X^2 $ (right plot).}\label{figKchar}
\end{figure}

Let us therefore focus on (\ref{char2}) for the moment. 
By substituting the solution of $\xi = \xi_{\pm}$ from (\ref{char}) and definitions of $A$ and $C$ into Eq.~(\ref{char2}), we obtain,
\begin{equation}\label{charG}
	\left(\frac{d\tau}{d\chi}\right)_{\pm} = -\xi_{\mp}.
\end{equation}
As one can see from (\ref{charG}), the characteristics $\Gamma_{\pm} \equiv \left(d\tau / d\chi \right)_{\pm} $ in the $(\tau,\chi)$ plane are 
connected to the characteristics $\xi_\pm$ in the $(t,x)$ plane in a particularly simple way,
$\Gamma_+ = -\xi_-$ and $\Gamma_- = -\xi_+$. 
It is worth to note that in the canonical case, $\mathcal{L}=X$, the characteristics are the straight lines $\xi_\pm^{\rm can}=\pm1$, independently on the value of $\chi$ and $\tau$,
since the equation of motion is linear. On the contrary, in the non-canonical case, the characteristics depend on the solution.
At this point it is convenient to introduce the following quantity, 
\begin{equation}\label{cs}
c_s^2 = \left(1+2X\frac{\mathcal{L}_{XX}}{\mathcal{L}_X}\right)^{-1}.
\end{equation}
Using (\ref{cs}), Eq.~(\ref{xipm}) can be simplified to,
\begin{equation}\label{xipm2}
	\xi_+= \frac{\tau c_s -\chi}{\tau - \chi c_s},\,\, 	\xi_-= - \frac{\tau c_s + \chi}{\tau + \chi c_s}.
\end{equation}
The quantity $c_s$ has the meaning of the speed of propagation of small perturbations with respect to the background solution with timelike $\partial_\mu\phi$. 
Indeed, from (\ref{xipm2}), for a background solution $\chi=0$, $\tau\neq 0$ (which, for example, corresponds to a homogeneous cosmological solution),
we get $\xi_\pm = \pm c_s$. Similarly, for static solutions, $\chi\neq 0$, $\tau = 0$, we obtain from (\ref{xipm2}) $\xi_\pm = \pm 1/c_s$.
In particular, for the linear theory, $\mathcal{L}(X) = X$, the sound speed is constant, $c_s=1$, as it can be seen from (\ref{cs}).
It is also convenient to introduce the ``velocity'' of the k-essence as follows,
\begin{equation}\label{defv}
	v =  -\frac{\chi}{\tau}.
\end{equation}
The minus sign in (\ref{defv}) is due to the difference of the co- and contravariant components of a vector. 
With the notations~(\ref{defv}), Eq.~(\ref{xipm2}) becomes simply the standard expressions for relativistic addition of velocities.
Using (\ref{defv}), the characteristic equations (\ref{xipm2}) can be integrated along each of the characteristics,
\begin{equation}\label{Riemann}
	h(X) + \ln\left(\frac{1+v}{1-v}\right) = C_1 (\sigma_-),\,\,\,
	h(X) - \ln\left(\frac{1+v}{1-v}\right) = C_2 (\sigma_+),
\end{equation}
where,
\begin{equation}\label{defh}
	h(X) = \int \frac{dX}{c_sX}.
\end{equation}
In the canonical case, Eq.~(\ref{defh}) can be easily integrated, and substituting the result in Eq.~(\ref{Riemann}), we obtain, 
\begin{equation}
\tau - \chi=\tilde C_1 (\sigma_-), \,\,\, \tau + \chi=\tilde C_2 (\sigma_+),
\end{equation}
from which, with the identifications $\sigma_\pm=t\pm x$, the standard result of the linear theory follows, 
\begin{equation}
	\phi(t,x) = \phi_1 (t-x) + \phi_2(t+x).
\end{equation}
 The characteristics for any solution are straight lines both in 
the real space time and the $(\tau,\chi)$ plane, as it immediately follows from (\ref{xipm2}), $\Gamma_\pm = \mp 1$.
The characteristic curves are not straight lines in a generic k-essence. 
In particular, 
Fig.~\ref{figKchar}  shows the characteristics for  $ \mathcal{L}(X) = X+ \frac12 X^2$ (left plot) and 
$\mathcal{L}(X) = X - \frac12 X^2$ (right plot) models. The model $ \mathcal{L}(X) = X+ \frac12 X^2$ corresponds to subluminal propagation of signals, as it can be inferred from (\ref{cs}) or (\ref{xipm2}). 
The grey regions in Fig.~\ref{figKchar} correspond to the values of $(\tau,\chi)$, where the equation of motion is not hyperbolic, for both models.
In the blue region of the left panel of Fig.~\ref{figKchar} (the model $ \mathcal{L}(X) = X+ \frac12 X^2$) the characteristics $\xi_\pm$ has the same sign, 
that means the signals only travel in one direction. 
The borders of the blue region corresponds to the ``acoustic'' horizons: either a white or a black hole. 
The situation with the model $ \mathcal{L}(X) = X - \frac12 X^2$ is quite different: any signal motion is luminal or 
superluminal\footnote{Superluminal propagation leads to interesting consequences, e.g. with the use of the k-essence signals it is possible to ``look'' inside the black hole horizon, see~\cite{Babichev:2006vx}.}. 
The blue colour on the right panel of Fig.~\ref{figKchar} (the model $ \mathcal{L}(X) = X- \frac12 X^2$), however, 
covers the region where the propagation along one of the characteristics is {\it backwards} in time, 
in contrast to the model $ \mathcal{L}(X) = X + \frac12 X^2$.

\section{Waves in k-essence}
\label{Section:Waves}
The characteristic method is especially convenient in the case of wave propagation in k-essence. 
In this section we consider particularly simple case of wave solutions, the so called {\it simple} waves.

\begin{figure}[t]
\includegraphics[width=0.85\textwidth]{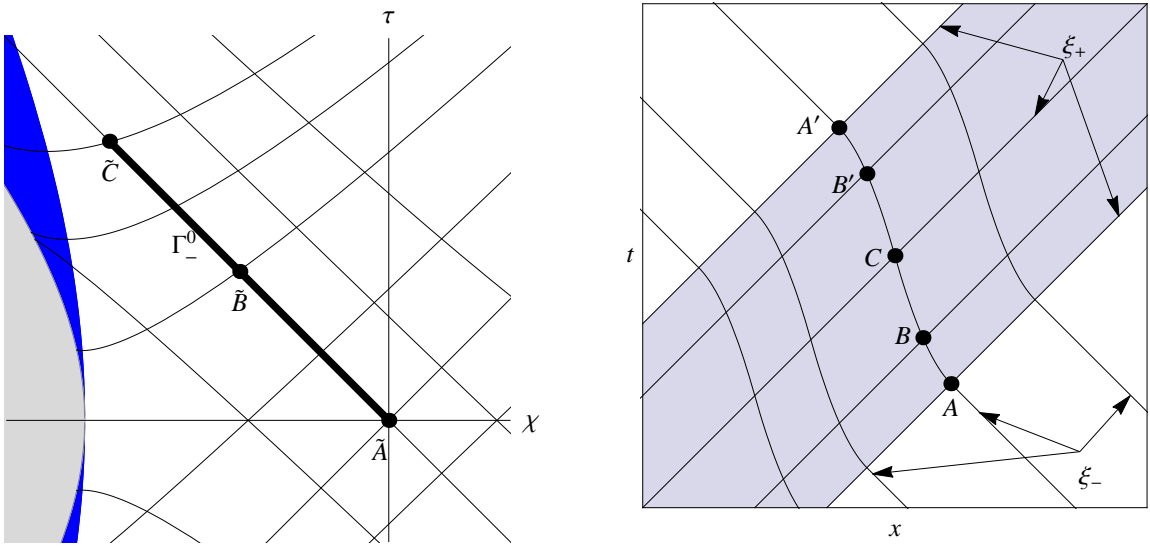}}{\caption{The solution of a wave propagating in the right direction is shown 
             for the model $\mathcal{L}(X) = X+ \frac12 X^2$. 
             This solution describes a ``travelling wave'', for which the shape of the wave does not change with time and no caustics form.
             On the left panel the travelling wave solution is shown in the $(\tau,\chi)$ plane. 
             The solution fully lies on the section $\tilde{A} \tilde{C}$ of a singe characteristic $\Gamma_-$, which is a straight line. 
             The same solution is shown on the right panel in the $(t,x)$ plane. 
		The right-directed characteristics $\xi_+$ are straight and {\it parallel} lines, as a consequence of the straight characteristic $\Gamma_-$ in the $(\tau,\chi)$ plane, 
	on which the image of the wave solution lies. The characteristics $\xi_+$ do not intersect each others. 
	The $\xi_-$ characteristics are not straight, but any $\xi_-$ is obtained from another $\xi_-$  by a shift $(t+\text{const},x+\text{const})$, therefore they do not intersect each 		others either.
             }\label{figtrav}
\end{figure}

A simple wave is a solution $\phi(t,x)$ of (\ref{eom0}), that the image of the solution in the $(\tau,\chi)$ plane fully lies on one characteristic, 
$\Gamma_+$ or $\Gamma_-$. A particular interesting physical situation when this happens is the following. In one region of spacetime $(t,x)$
the solution is stationary, i.e. $\dot\phi$ and $\phi'$ are constants. 
The simplest case of stationary solution is $\phi=\text{const}$, whilst a slightly more complicated one is 
$\phi = C t$, where $C$ is some constant,
the latter can be motivated by cosmological models involving a scalar field.  
Adjacent to this region, consider another region of spacetime $(t,x)$, where 
the solution is non stationary, i.e. $\dot\phi$ and $\phi'$ are not constants. 
This non-stationary (non trivial) solution is a propagating wave. The fact that it is a simple wave can be shown as follows, see Fig.~\ref{figtrav}. 
The region with a stationary solution (while color in the right panel of Fig.~\ref{figtrav}) is separated from the wave region (grey) by a characteristic line, in this case by $\xi_+$. 
The family of $\xi_-$ characteristics fall in the same characteristic $\Gamma_-$ in the $(\tau,\chi)$ plane, left panel of Fig.~\ref{figtrav}. 
This property holds because the image of all $\xi_-$ characteristics in the stationary region is one point $\tilde{A}$ in the $(\tau,\chi)$ plane, since 
in the stationary region  $\tau$ and $\chi$ are constants. At the same time, the image of any $\xi_+$ characteristic is a dot in the $(\tau,\chi)$ plane, 
e.g. $\tilde{A}$, $\tilde{B}$, $\tilde{C}$ in Fig.~\ref{figtrav}.
By continuity, the images of the characteristics $\xi_-$ coincide in the wave region, the thick line in the left panel of Fig.~\ref{figtrav}, 
and therefore by definition such a solution is a simple wave. 
Below we consider wave  propagation in k-essence in  detail. 

The characteristics $\xi_+$ passing through the points $A$ and $A'$ in Fig.~\ref{figtrav} divide the propagating wave from the static (or stationary) state.
The solution can have discontinuous second derivatives across these characteristics, and equations of motion are still well defined. 
This is in contrast to the case of formation of caustics we discuss below, when the first derivatives are also discontinuous at the point of caustics formation.

\subsection{Travelling wave}
\label{ssecTW}
We first  show that the solution found in~\cite{Babichev:2007dw} for propagating k-essence waves fits
a special case of a general simple wave. 
Notice that the characteristics $\Gamma_\pm$  
passing through $\tau=\chi=0$ in the $(\tau,\chi)$ plane are straight lines, $\Gamma_\pm=\pm 1$.
This can be immediately seen from (\ref{xipm}) by setting $X=0$ and $\tau=\pm\chi$ for $\xi_\pm$ correspondingly (remember also that
$\Gamma_+ = -\xi_-$ and  $\Gamma_- = -\xi_+$).
This is related to the Lorenz invariance of the considered theory.
This property is crucial for the solutions found in Ref.~\cite{Babichev:2007dw} to exist.
Indeed, let us consider a wave, such that the image of its characteristics lies on $\Gamma_-^0$, passing through $\tau=\chi=0$ 
(we can also take $\Gamma_+$, of course,
but for definiteness we concentrate on the case $\Gamma_-$), the point $\tilde{A}$ in Fig.~\ref{figtrav}.
Since $\Gamma_-^0=-1$, we have,
\begin{equation}
\tau+\chi =0 \Leftrightarrow  \dot\phi+\phi'=0
\end{equation}
which leads to the wave solution,
\begin{equation}\label{soltw}
\phi = \phi(t-x)
\end{equation}
This is exactly the solution presented in~\cite{Babichev:2007dw}.
Clearly, the other travelling wave also exist, 
\begin{equation}\label{soltw2}
\phi = \phi(t+x).
\end{equation}
However a combination of the two is a solution only for the canonical scalar, $\mathcal{L}=X$, 
while for a nonlinear Lagrangian a sum of  $\phi(t-x)$ and $\phi(t+x)$ is not a solution.

An interesting property of this travelling wave solution is that  
the $\xi_+$ characteristics in $(t,x)$ plane are straight lines with $\xi_+=1$ everywhere, 
due to the relation $\Gamma_- = -\xi_+$, see also Fig.~\ref{figtrav}.
Note that in general $\xi_-$ are not straight lines (except for the linear theory), since $\xi_-=-\Gamma_+$, and 
the value of $\Gamma_+$ depends on the position in $\Gamma_-^0$. 
The fact that $\xi_+=1$ everywhere means that the wave keeps its form while propagating 
(which is also clear from the form of the solution for the travelling wave~(\ref{soltw})).
Another important consequence is that the speed of perturbations 
in the direction of wave propagation {\it is constant and it equals to the speed of light}, $c_s=1$.

\begin{figure}[t]
\includegraphics[width=0.85\textwidth]{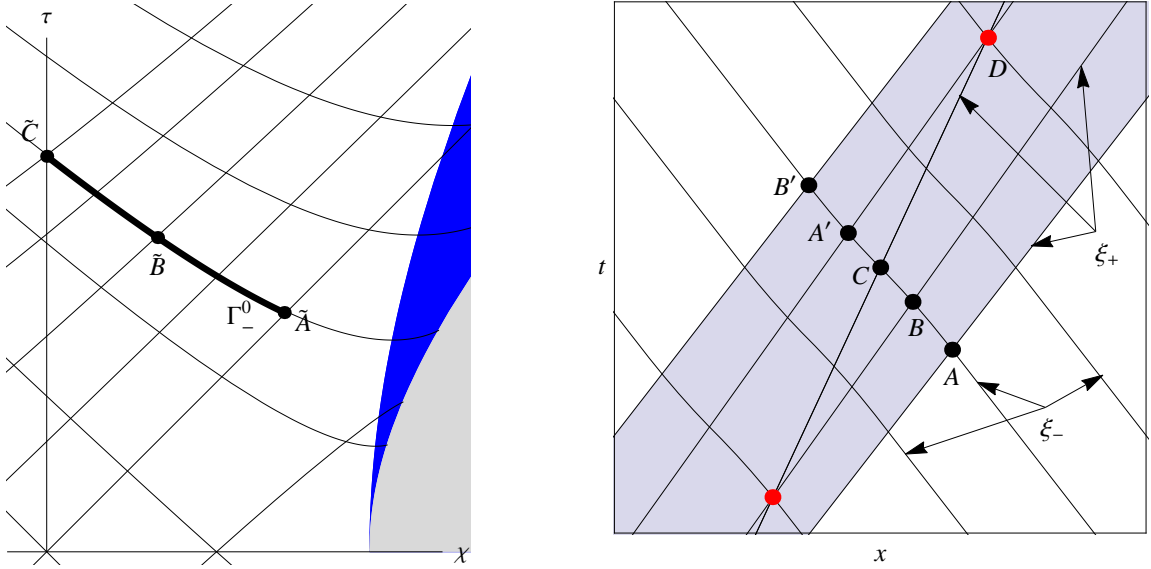}}{\caption{
	The solution of a wave propagating in the right direction is shown 
             for the model $\mathcal{L}(X) = X+ \frac12 X^2 $ in the $(\tau,\chi)$ plane (left panel) and in the $(t,x)$ plane (right panel). 
             In the $(\tau,\chi)$ plane the solution fully lies on the section $\tilde{A}\tilde{B}\tilde{C}$ of a singe characteristic $\Gamma_-^0$ . 
             The characteristic $\Gamma_-^0$ is not a straight line in this case (compare to Fig.~\ref{figtrav})
             This solution leads to formation of caustics, as it is shown on the right panel. 
             The right-directed characteristics $\xi_+$ {\it are not parallel}, unlike the solution in Fig.~\ref{figtrav} and caustics form, 
	when different characteristics intersect (shown by red dots.)
             }\label{figsimple}
\end{figure}

It is interesting to mention that for a theory $\mathcal{L}=\mathcal{L}(\phi,X)$ in the case of initial data close to zero, 
one can formulate a criterion for the smooth solution to exist globally~\cite{Rendall:2005fv}. 
It happens that for pure k-essence, this condition is satisfied automatically. 
Assuming small $X$, travelling wave solutions may be considered as a particular class of such solutions. On the other hand,  more general wave solutions, studied below in Sec.~\ref{ssecGW}, 
do not have small $X$ limit, thus the above condition is violated.

Note also that in a class of k-essence theories with explicit dependence on $\phi$, $\mathcal{L}=\mathcal{L}(\phi,X)$, for which static soliton solutions can be constructed~\cite{Babichev:2006cy}, 
thanks to symmetry-breaking. In this case any boost of the found solution will formally give a ``travelling wave'', however, this type of solutions have a different physical origin.

\subsection{Generic waves in k-essence. Formation of caustics}
\label{ssecGW}
The situation described above is a particular case of wave propagation, which requires initial conditions $X=0$ as we have seen in the previous section~\ref{ssecTW}.
The general wave solution violates this condition. 
In fact, a physical situation leading to $X\neq 0$ is quite generic. In particular, a wave propagating on a cosmological background has $X\neq 0$.
Indeed,
for a homogeneous cosmology we have $\chi = \phi' =0$, while $\dot\phi$ is generically non-zero, $\tau \neq 0$ 
(in k-essence models designed to explain the present-day acceleration of the Universe, $\dot\phi$ is necessarily non-zero). 
For simplicity assume a constant  $\dot\phi$,  i.e. $\tau_c=$const. This assumption does not affect the main result. 
Consider a propagating wave travelling in the positive $x$ direction. 
In the $(\tau,\chi)$ plane, the whole wave lies on the characteristic $\Gamma^0_-$, see Fig.~\ref{figsimple}, 
similar to the case considered in section~\ref{ssecTW}.
The homogeneous cosmology corresponds to the point $\tilde{C}$ on the left panel and to the white regions on the right panel of Fig.~\ref{figtrav}.
The wave is shown by the grey color and its image lies on the characteristic $\Gamma^0_-$. 

The characteristics $\xi_+$ are straight lines in the $(t,x)$ domain, because each characteristic $\xi_+$ collapses to a point in the $(\tau,\chi)$ plane. 
However, in contrast to the travelling wave,
the slope of a characteristic $\xi_+$ now depends on the position, see the right panel of Fig.~\ref{figsimple}, 
therefore the characteristics $\xi_+$, although being straight lines, are not parallel. 
In particular, in $\xi_+$ has different values at the points $A$, $B$, $C$. 
This is a consequence of the fact that $\Gamma_-^0$ is not a straight line, see the left panel of Fig.~\ref{figsimple}.
In particular, $\Gamma_-$ at the points $\tilde{A}$, $\tilde{B}$ and $\tilde{C}$ has different values.
This has a drastic effect: {\it characteristics intersect} (shown by red dots in Fig.~\ref{figsimple}), 
which means {\it the appearance of caustics}. At the points of caustic formation, the values of $\tau$ and $\chi$ are not single-valued, 
since, two or more characteristics, carrying the same values of $\tau$ and $\chi$ intersect at one point of space-time. 
In the left panel of Fig.~\ref{figsimple} the 
two characteristics, passing through $A'$ and $C$ intersect in the future, at point $D$. 
The values of $\tau$ and $\chi$ are constants along each of the $\xi_+$ characteristics, but $\xi(A) \neq \xi(C)$.

At the intersection of characteristics (some of) the second derivatives of $\phi$ diverge. 
Indeed, when approaching the point $D$, the distance $\delta x$ between characteristics tends to zero, 
while the difference between $\chi$ at different characteristics stays the same, therefore, for example $\phi'' \sim \delta \chi/\delta x \to \infty$.
When a caustic forms, Eq.~(\ref{eom0}) becomes singular and the theory loses predictability. 

Thus the caustic formation appears to be a generic feature of the k-essence theories. 
This can also be seen from the study of the initial data problem.
Let $t=0$ be the initial data hypersurface, for definiteness. 
In general, according to the Cauchy problem, two initial conditions must be specified, $\phi(0,x)=\phi_0(x)$ and $\dot\phi(0,x) = \phi_1(x)$, 
where $\phi_0$ and $\phi_1$ are arbitrary smooth functions. In the case of a simple wave, the initial conditions are restricted, 
since, by definition, an image of a simple wave solution fully lies on one characteristic in the $(\tau,\chi)$ plane. 
In fact, as it will be clear in a moment, only one of the two initial conditions must be imposed, while the second condition 
is a consequence of the restriction that we look for is a simple wave. 
Indeed, having specified the value of $\phi$ on the hypersurface $t=0$, $\phi(0,x)=\phi_0(x)$, one fixes $\chi = \phi'_0(x)$ at the hypersurface.
This immediately implies that $\tau = \dot\phi $ is also fixed by the definition of a simple wave, 
see~Fig.~\ref{figsimple}, and therefore for a simple wave each value of $\chi$  uniquely defines $\tau$.
In particular, for any simple wave solution (for definiteness, we consider the right-propagating wave), 
at the hypersurface of initial data, $t=0$, the value of the characteristic $\xi_+(0,x)$ can be expanded around the particular $x=0$ as,
\begin{equation}\label{xistr}
	\xi_+(0,x) = \xi_+(0,0) + \beta x + \mathcal{O}(x^2).
\end{equation}
The propagation of the characteristics $\xi_+$ is given by straight lines. Then from~(\ref{xistr}) one can deduce that two characteristics, starting from two different points $x_1$ and $x_2$ at $t=0$, 
will collide after certain time, given by
\begin{equation}
t=-\frac{1}{\beta},
\end{equation}
where one assumes that $\beta<0$ (for $\beta>0$ the collision ``happens'' in the past). For $\beta =0$, the characteristics do not collide, this case was studied in (\ref{ssecTW}).

It can be seen now that the solution of considered in the previous subsection (\ref{ssecTW}) (see also~\cite{Babichev:2007dw}), 
corresponds to a special choice of initial conditions. More precisely, a travelling wave of Sec.~(\ref{ssecTW}) 
corresponds to a specific choice of the characteristic, which defines the simple wave, namely a characteristic satisfying $X=0$. 
For all other simple waves, for which $X\neq 0$, the wave solution changes its shape when propagating, which eventually leads to the caustic formation. 

\section{Waves in Galileon}

Before extending our study of wave propagation to include higher-order Lagrangians, let us first notice an important property of the wave solutions we considered so far in Sections~\ref{ssecTW} and \ref{ssecGW}.
A generic simple wave solution satisfy the following relation,
\begin{equation}\label{jacobian}
	\ddot\phi \phi'' - (\dot\phi')^2 =0.
\end{equation}
It is easy to verify (\ref{jacobian}) for the travelling wave of the section \ref{ssecTW}, due to the simple form of the corresponding solutions, $\phi(t,x) = f(t-x)$ or $\phi(t,x) = f(t+x)$.
For the general wave solution considered in section \ref{ssecGW}, we need to invoke the definition of the general simple wave: 
it is a solution whose image completely lies on one characteristic in the $(\tau,\chi)$ plane. 
For example, the image of the solution shown in Fig.~\ref{figsimple}, lies on a certain $\Gamma_-^0$ characteristic, 
the one which passes through the points $\tilde{A}$, $\tilde{B}$ and $\tilde{C}$.
Using this feature, it is not difficult to get the following chains of relations, 
\begin{equation}\label{jacbis}
	\ddot\phi = \dot\tau=  \frac{d\tau}{d\sigma_-}\dot\sigma_-, \,\,\,
	\phi'' = \chi'=  \frac{d\chi}{d\sigma_-}\sigma_-', \,\,\,
	\dot\phi' =  \frac{d\tau}{d\sigma_-}\sigma_-' = \frac{d\chi}{d\sigma_-}\dot\sigma_-.
\end{equation}
Combining (\ref{jacbis}) we arrive at (\ref{jacobian}). This relation, (\ref{jacobian}), 
will be crucial for what follows in the rest of this subsection. 
On the other hand, Eq.~(\ref{jacobian}) is exactly 
the condition which does not allow to perform a {\it hodograph} transformation, and exchange the dependent and independent variables~\cite{Courant}.

As we saw in section~\ref{ssecGW}, 
formation of caustics is rather generic in k-essence. 
In order to ``smooth out'' caustics, we can try to modify the action (\ref{L}) of the theory, by adding extra terms.
However, we would like to naturally impose the following restrictions for modifications:
i) the resulting equation is of the second order in time, so that the Ostrogradski ghost does not appear; and 
ii) the Lorentz invariance is preserved. 
Such theory has been formulated by Horndeski~\cite{Horndeski} and later was 
re-descovered in~\cite{Deffayet:2011gz} by a generalisation of the ``Galileon''~\cite{Nicolis:2008in,Deffayet:2009wt} (The equivalence between the Galileon and Horndeski theory has been established in~\cite{Kobayashi:2011nu}). 
We need only a part of the Horndeski theory, since we assume that the metric is non-dynamical. 
Also we restrict our study to the shift-symmetric case, as we did in the case of pure k-essence.

A convenient way to write the galileon action is to use the fully antisymmetric tensor $\epsilon_{\mu\nu\rho\sigma}$~\cite{Deffayet:2009mn}. 
The galileon Lagrangian can be then written as a sum of the different terms,
\begin{equation}
\begin{split}\label{action g}
	\mathcal{L}_3 &= g_3(X) \epsilon^{\mu\nu\rho\sigma}\epsilon_{\alpha\beta\rho\sigma}\nabla_\mu\phi\nabla_\alpha\phi(\nabla_\nu\nabla^\beta\phi),\\
	\mathcal{L}_4 &= g_4(X) \epsilon^{\mu\nu\rho\sigma}\epsilon_{\alpha\beta\gamma\sigma}\nabla_\mu\phi\nabla_\alpha\phi(\nabla_\nu\nabla^\beta\phi)
		(\nabla_\rho\nabla^\gamma\phi),\\
	\mathcal{L}_5 &= g_5(X) \epsilon^{\mu\nu\rho\sigma}\epsilon_{\alpha\beta\gamma\delta}\nabla_\mu\phi\nabla_\alpha\phi(\nabla_\nu\nabla^\beta\phi)
		(\nabla_\rho\nabla^\gamma\phi)(\nabla_\sigma\nabla^\delta\phi),
\end{split}
\end{equation}
where $g_i(X)$ are arbitrary functions of the kinetic term.
Notice that in the $1+1$ dimensional motion the terms $\mathcal{L}_4$ and $\mathcal{L}_5$ do not contribute to the equations of motion, 
because of the antisymmetric nature of the epsilon-tensor: 
$\epsilon^{\mu\nu\rho\sigma}\epsilon_{\alpha\beta\gamma\sigma}$ and $\epsilon^{\mu\nu\rho\sigma}\epsilon_{\alpha\beta\gamma\delta} $ are zero in the case of two dimension. 
Therefore  the only non-trivial part of the full Horndeski model in the $1+1$ dimensional motion is given by the Lagrangian $\mathcal{L}_3$ of (\ref{action g}), which can be rewritten as
\begin{equation}\label{action dgp}
	\mathcal{L}_{3} = G(X)\Box\phi,
\end{equation}
where $G(X)$ is an arbitrary function of $X$.
The variation of the action $S_3 = \int d^4 x \mathcal{L}_{3} $ gives,
\begin{equation}\label{eom3}
	\frac{\delta S_3}{\delta \phi} = G_X(X) \left( (\Box\phi)^2 - \nabla_\mu\nabla_\nu\phi\nabla^\mu\nabla^\nu\phi\right)
	+ G_{XX}\left(\Box\phi\nabla_\mu\phi\nabla^\mu X -\nabla_\mu X\nabla^\mu X \right) =0.
\end{equation}
It is easy to see that for an ansatz $\phi=f(x-t)$ or $\phi=f(x+t)$ Eq.~(\ref{eom3}) is automatically satisfied, 
therefore a $\phi=f(x\pm t)$ is a solution for the theory $\sim X + X\Box\phi  $, in accordance with~\cite{Evslin:2011vh}. 
Moreover, it can be shown the travelling wave $\phi=f(x\pm t)$ is s solution for the full Horndeski theory as well~\cite{Babichev:2012qs}.

Let us see how the presence of the galileon Lagrangians affect the general wave solution we studied in section~\ref{ssecGW}.
For the general case $\phi=\phi(t,x)$, Eq.~(\ref{eom3}) gives, in two dimensions, 
\begin{equation}
	\frac{\delta S_3}{\delta \phi} = -2 G_X\left( \ddot\phi\phi'' - \dot\phi'^2 \right)  - 2 X G_{XX} \left( \ddot\phi\phi'' - \dot\phi'^2 \right).
\end{equation}
It is clear that for $\phi(t,x)$ for which (\ref{jacobian}) holds, the equation of motion for the galileon is identically zero. 
This means, that the general wave solution which we studied in section \ref{ssecGW} is also a solution for the generalised galileon theory,
\begin{equation}
\label{Lgal}
	\mathcal{L} = \mathcal{K}(X) + \mathcal{L}_3+ \mathcal{L}_4 + \mathcal{L}_5.
\end{equation}
As a consequence of this fact, caustics are a generic feature of the galileon theory as well. 

\section{Conclusions}

In this paper we studied in detail propagation of waves in $1+1$ dimensions in the k-essence and the galileon models by the methods of characteristics. 
We found the following main results:
\begin{itemize}
\item 
We confirmed that for the k-essence and the galileon model the travelling wave solutions~(\ref{soltw}) and (\ref{soltw2}) exist, in accordance to previous results in the literature. 
However, as we have shown in Sec.~\ref{Section:Waves}, this travelling wave solutions correspond to fine-tuned initial conditions. 
\item 
The general initial conditions for the k-essence model lead to a propagating wave, which does not correspond to a travelling wave. 
The propagating wave does not preserve its form along its evolution. 
Eventually for a generic wave the evolution leads to formation of caustics. 
After a caustic is formed, a solution of the differential equation cannot be continued.
\item 
Any wave solution for the k-essence model~(\ref{L}) is also a solution for the most general shift-symmetric galileon theory~(\ref{Lgal}), 
where the k-essence part of the galileon model is the same as in~(\ref{L}).
\end{itemize}

The above results suggest that both k-essence and the generalised galileon models cannot be considered as fundamental. 
At the moment when the caustics form, one should give extra input about the theory, which would allow to deal with the solutions at and after formation of caustics.
This is similar to the scenario when dust is effectively considered as continuous fluid with the zero speed of sound. 
When the particle of dust intersect (which correspond to formation of caustics in the fluid approximation), one should abandon the effective fluid description and consider each particle separately. In the case of the generalised galileon (and, as a particular subclass --- k-essence), the model 
should be completed by a some underlying theory, which would allow to resolve the caustic problem
(see Appendix for a tentative approach in this direction).

We would like to note that although the study in this paper has been restricted to the case of 1+1 dimensional spacetime, 
the results we find apply for the 4D spacetime by extending 2D solutions to 4D with the identification $\phi_{4D}(t,x,y,z) = \phi(t,x)$. 

It should be also stressed that although our results imply that in a generic galileon model there is caustics form, 
there is a special case of a galileon, for which the problem caustics is avoided in $1+1$ dimensional motion. 
Indeed, caustics form when a non-linear $K(X)$ term is present. 
But, on the other hand, if the k-essence term is canonical, $K(X)\sim X$ 
then caustics do not form (in 1+1), independently on the form of other (higher-derivative) galileon terms.
This may suggest that galileon models with $K(X)\sim X$ are healthy. 
To address this question, however, a study of galileon dynamics in 3D and 4D is required.

In our study we assumed shift-symmetric models. What happens if the dependence on the scalar field itself is allowed? 
Because the problem of caustics happens when the field develops large derivatives (even infinite), 
it is safe to assume that the dependence on $\phi$ does not affect the main results.

One possible way to avoid the problem of caustics would be to make the  metric dynamical. 
Since at the points where caustics form, the first derivatives of $\phi$ experience a jump, 
the backreaction of the gravitational field may be important at those points, therefore the dynamics of the scalar field must be modified accordingly. 
Thus the gravitational backreaction may in principle prevent formation of caustics. This question, however, lies beyond the scope of this paper.

Finally, one may adopt a more phenomenological point of view, and provide a practical prescription for treating the galileon model at the location 
of caustics.
In this approach there is no need to resort to an underlying theory for galileon. 
In particular, one may think of shock waves appearing once caustics form. In this case an extra input is needed to describe propagation of shock waves in galileon, 
such that it would not conflict with conservation of the energy-momentum tensor.
This approach, in fact, may open interesting possibilities for phenomenology of scalar-tensor models: 
in particular, shock waves in k-essence and galileon may affect the inflation spectrum, and 
change the picture of reheating at the end of inflation. 

\vspace{.15in}

\noindent
{\bf Acknowledgments}\quad I would like to thank Alexander Lychagin for inspiring discussions and 
Gilles Esposito-Far\`ese for useful discussions and careful reading of the manuscript. 
This work was supported in part by Russian Foundation for Basic Research Grant No. RFBR 15-02-05038.

\appendix
\section{K-essence as a limit of nonlinear sigma model}
Let us consider the following Lagrangian,
\begin{equation}\label{Lsigma}
	\tilde{\mathcal{L}}= \frac{\varepsilon}2(\partial\lambda)^2 + \frac{\lambda}2 (\partial\phi)^2 - V(\lambda),
\end{equation}
where $\varepsilon$ is a constant and $\phi$ and $\lambda$ are the two propagating degrees of freedom of the theory. 
The above Lagrangian (\ref{Lsigma}) is a typical example of a non-linear sigma model. 
Variation of the action $S=\int d^4 x\tilde{\mathcal{L}}$ with respect to $\lambda$ and $\phi$ yields, correspondingly,
\begin{eqnarray}
	-\varepsilon \Box\lambda + \frac12(\partial\phi)^2 - V'(\lambda) &=& 0 \label{eomsigma1}, \\
	-\nabla^\mu\left( \lambda\partial_\mu\phi \right) &=&0.\label{eomsigma2}
\end{eqnarray}

If we take $\varepsilon =0$ in the above equations, one can resolve~(\ref{eomsigma1}) in order to find $\lambda$ in terms of $(\partial\phi)^2$, i.e.
\begin{equation}\label{lambdaf}
\lambda = f[(\partial\phi)^2],
\end{equation}
where $f$ is inverse of $V'$. Substituting the obtained expression in (\ref{eomsigma2}), one obtains the k-essence EOM (\ref{eom0}), 
with the identification $f \equiv \mathcal{L}_X$ ($\mathcal{L}$ here is a pure k-essence Lagrangian as in (\ref{L})).
Therefore, for $\varepsilon =0$ the model (\ref{Lsigma}) is merely another form of the same pure k-essence theory~(\ref{L}).
For $\varepsilon \neq 0$ the two theories are clearly different, in particular because (\ref{Lsigma}) contains two propagating degrees of freedom, while (\ref{L}) has only one.

Now, let us assume that $\varepsilon$ is small but nonvanishing. 
For any smooth solution and for small enough $\varepsilon$, Eq.~(\ref{lambdaf}) is a good approximation to an exact solution of (\ref{eomsigma1}).
Therefore, with this approximation, we recover the equations of motion for k-essence and the non-linear sigma model effectively reduces to k-essence. 
Of course, this is true only if $\varepsilon\Box\lambda$ 
is negligible in comparison to other terms in (\ref{eomsigma1}). 
Once the first term in this equation becomes important, the sigma-model ceases to correctly describe k-essence.
It is important to stress that the principal part of Eqs.~(\ref{eomsigma1}) and (\ref{eomsigma2}) has (almost) canonical structure,
therefore one expects that the characteristics for the sigma-model to be light-like (at least this is true for small perturbations on a non-trivial background in the high-frequency limit).
This implies that no caustics should form in the sigma-model. Thus the model~(\ref{Lsigma}) may be a good candidate for a (classical) completion of the k-essence theory.
We may expect that sigma-model completion resolves caustic formation, giving instead smoothed shock waves, with everywhere well-defined equations of motion.
This question will be studied elsewhere.


\end{document}